\documentclass{aa}  
\usepackage{xcolor}
\usepackage{ulem}
\usepackage{graphicx}
\usepackage{multirow}
\usepackage{txfonts}

\begin{document}

   \title{Density constraint of the warm absorber in NGC 5548}

   \author{Keqin Zhao
          \inst{1}, 
          Jelle S. Kaastra\inst{1}\fnmsep\inst{2}, 
          \and
          Liyi Gu\inst{2}
          }

   \institute{Leiden Observatory, Leiden University, PO Box 9513, 2300 RA Leiden, The Netherlands\\
              \email{kzhao@strw.leidenuniv.nl}
         \and
             SRON Space Research Organization Netherlands, Niels Bohrweg 4, 2333 CA Leiden, The Netherlands\\
             }

   \date{Received January 14, 2025; accepted September 17, 2025}

  \abstract
   {Ionized outflows in active galactic nuclei (AGNs) are thought to influence the evolution of their host galaxies and supermassive black holes (SMBHs). Distance is important to understand the kinetic power of the outflows as a cosmic feedback channel. However, the distance of the outflows with respect to the central engine is poorly constrained. The density of the outflows is an essential parameter for estimating the distance of the outflows. NGC 5548 exhibits a variety of spectroscopic features in its archival spectra, which can be used for density analysis.}
   {We aim to use the variability of the absorption lines from the archival spectra to obtain a density constraint and then estimate the distance of the outflows.}
   {We used the archival observations of NGC 5548 taken with Chandra in January 2002 to search for variations of the absorption lines.}
   {We found that the \ion{Mg}{XII} Ly$\alpha$ and the \ion{O}{VIII} Ly$\beta$ absorption lines have significant variation on the $\sim$144 ks time scale and the $\sim$162 ks time scale during the different observation periods. Based on the variability timescales and the physical properties of the variable components that dominated these two absorption lines, we derive a lower limit on the density of the variable warm absorber components in the range of $\sim$7.2-9.0$\times10^{11}\text{m}^{-3}$, and an upper limit on their distance from the central source in the range of $\sim$0.2-0.5 pc.}
   {} 

   \maketitle

\section{Introduction}
Ionized outflows in AGNs which transport matter and energy away from the nucleus, thus linking the SMBHs to their host galaxies, are thought to influence their nuclear and local galactic environment (\citealp[e.g.][]{1998A&Afeedback1, feedback22010MNRAS, feedback32006MNRAS, feedback42001}). X-ray observations are crucial to characterizing these outflows. Through the application of medium- and high-resolution X-ray spectroscopy, researchers have discovered that these outflows are characterized by absorption lines of photoionized species that appear blue-shifted with respect to the systemic velocity of the host galaxies (\citealt{intro2000A&AJelle}). These blue-shifted absorption features provide direct evidence of the outward motion of the gas and allow quantitative measurements of outflow properties. Warm absorbers (WAs) are one of the manifestations of photoionized outflows, which typically exhibit hydrogen column densities ranging from 10$^{24}$ to 10$^{26}\text{m}^{-2}$ and outflow velocities of approximately $\sim 100-1000\ \text{km\ s}^{-1}$ (\citealp{WAparameters11997MNRAS, WAparameters22005A&A, WAparameters32012A&A, WAparameters42021}). For detailed investigations of these outflow phenomena, the nearby and bright Seyfert AGNs provide the best laboratories due to their proximity and luminosity, allowing us to obtain high-quality spectra with sufficient signal-to-noise ratios to characterize the complex absorption features.

The origin and launching mechanism of the ionized outflows in AGN remain uncertain. To better understand these mechanisms, determining the location of these outflows relative to the central source is crucial, as this can help distinguish between different theoretical models. The distances can be indirectly constrained by measurements of the ionization parameter, ionizing luminosity, and density. Although the ionization parameter and ionizing luminosity can be obtained from spectral fitting, accurately assessing the density of the outflow remains a challenge. An effective approach is timing analysis, where the response of the ionized outflow to changes in the ionizing continuum is monitored over time. Ideally, tight limits can be placed on the distance by measuring the variability in the ionization properties of the warm absorber in response to changes in incident ionizing flux (\citealp[e.g.][]{Method12008A&A, Method22010A&A, Method32012A&A}).

The archetypal Seyfert-1 galaxy NGC 5548 is one of the most widely studied nearby active galaxies both in the X-rays (\citealp{Jelle2002A&Adatareduction, steenbruggelinesearch2005A&A, 5548relatedD2008A&A...488...67D, 5548relatedD2009A&A, 5548related42010ApJ, 5548related52010ApJ}) and in the ultraviolet (\citealp{UV11999ApJ, UV22002ApJ, UV3, UV42003ApJ, UV52009ApJ}). The available XMM-Newton and Chandra grating data of this object have accumulated to >2 Ms in total, making it one of the deepest spectroscopic AGN datasets so far and the primary spectral components have been modeled to good precision in previous works (\citealt{Liyi2022A&A}). These advantages make NGC 5548 as an appropriate target to allow us to measure the variability.
\begin{table}
\centering
\caption{The exposure time and details for the Chandra observations of NGC 5548 used in this paper.}
\begin{tabular}{llccl}
\hline\hline 
 Instrument& Start time & Exposure time (ks) & Label \\ \hline
 HETGS& 2002 Jan. 16 & 152.1&MEG  \\
 LETGS& 2002 Jan. 18 & 59.0 &LETGSa\\
 LETGS& 2002 Jan. 18 & 84.0 &LETGSb\\
 LETGS& 2002 Jan. 21 & 84.9 &LETGSc\\
 LETGS& 2002 Jan. 21 &113.0&LETGSd\\ \hline
\end{tabular}
\label{tab:spectra}
\end{table}   

In this work, we constrain the lower limit density range of the WAs in NGC 5548 by reanalyzing the 2002 Chandra High Energy Transmission Grating Spectrometer (HETGS) and low-energy transmission grating spectrometer (LETGS) data, based on the variability of absorption lines and the accumulated knowledge on the primary spectral components. This paper is organized as follows. Section~\ref{sec:Observation and data reduction} provides all the observations used and the detailed data reduction. Section~\ref{sec:Data analysis} gives the characteristics of the light curves and the absorption lines variation and shows the results of the density constraint. Finally, we compare our results with the previous work and estimate the distance in Section~\ref{sec:discussion}.

\section{Observation and data reduction}
         \label{sec:Observation and data reduction}

\begin{figure*}
\centering
\includegraphics[width=\textwidth]{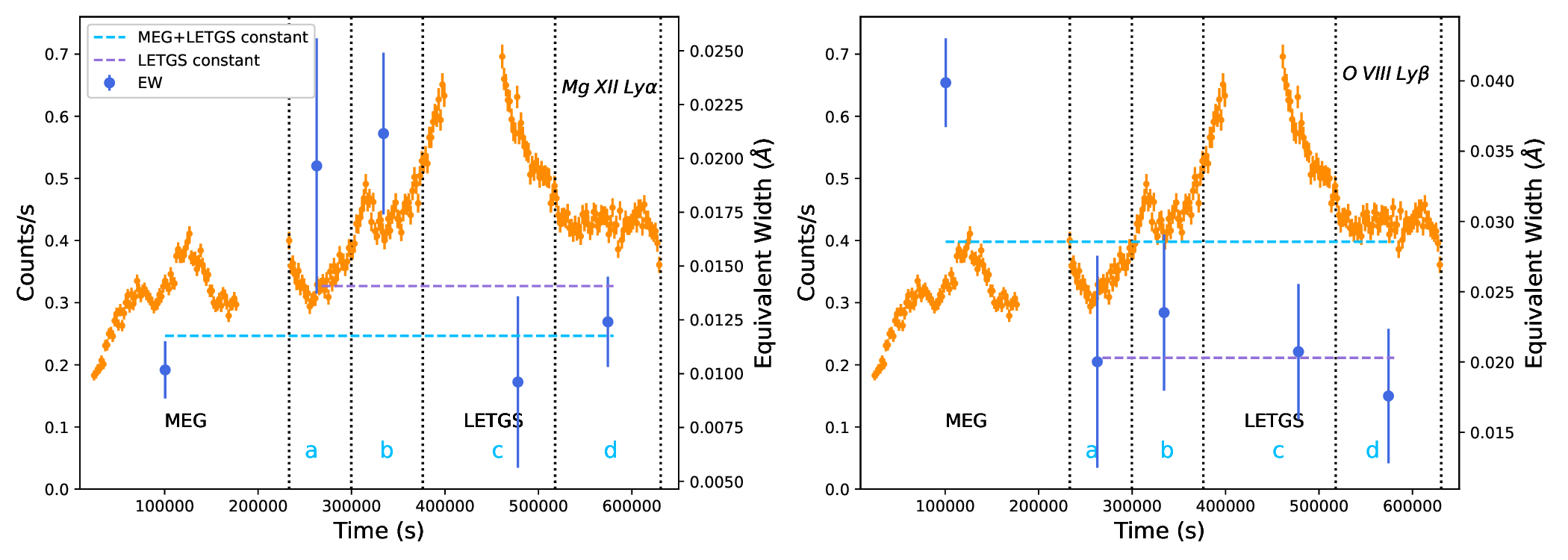}
\caption{The equivalent width variation over time. The light curve during the MEG and LETGS observations from \cite{Jellelightcurve2004A&A} is plotted in the background as orange points. The vertical gray dotted lines and blue lowercase letters mark the split LETGS spectra discussed in the text. The blue points are the equivalent width values. The dashed lines represent the average values from Table~\ref{tab:EWfitting}: light blue lines for the MEG and LETGS observations, purple lines for the LETGS observation only.}
\label{fig:EWfit}
\end{figure*}

\begin{figure*}
\centering

\includegraphics[width=\textwidth]{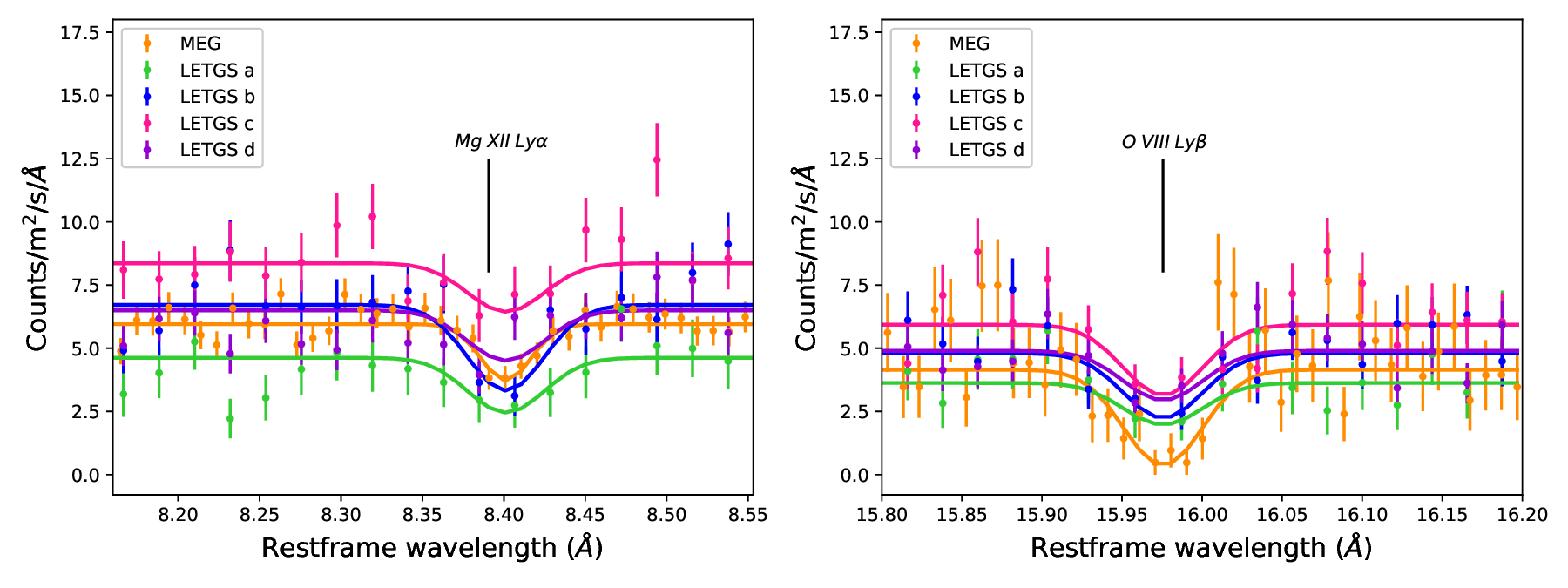}
\caption{The line profiles of the \ion{Mg}{XII} Ly$\alpha$ and \ion{O}{VIII} Ly$\beta$ lines with the best-fit Gaussian profile for the MEG and LETGS spectra.}
\label{fig:lineprofile}
\end{figure*}

The archival Chandra observation of NGC 5548 sums up to $\sim500$ ks of exposure of grating spectra in January 2002, including both the HETGS and LETGS data. The HETGS data were reduced using the standard CIAO software version 2.2. The detailed LETGS data reduction has been described in \cite{Jelle2002A&Adatareduction}. The HETGS spectra consist of a High Energy Grating (HEG) spectrum and a Medium Energy Grating (MEG) spectrum. We searched the line features in the $1.5-38$ $\AA$ range with a FWHM of 0.05 $\AA$ for the LETGS spectra and in the $1.5-24$ $\AA$ range with a FWHM of 0.023 $\AA$ range for the MEG spectra. All spectra are binned to 0.5 FWHM and are the same as used by \cite{steenbruggelinesearch2005A&A}. The LETGS observation was split over two orbits of the Chandra satellite (170 ks exposure, starting January 18, and 171 ks exposure, starting January 21, respectively). The aim of this work is to search for variability in individual absorption lines. In order to find the variation features and keep a good signal-to-noise ratio in the meantime, we split all LETGS observations into four pieces. In what follows, the MEG and split LETGS spectra will be identified in the paper as MEG, LETGSa, LETGSb, LETGSc, and LETGSd, respectively. In Table~\ref{tab:spectra} the instrumental setup and exposure times are listed.

\section{Data analysis}
\label{sec:Data analysis}
\subsection{Light curve}
We investigated the response of the ionized outflow to changes in the ionizing continuum. Therefore, the continuum flux variation is important. The light curve extracted from the zeroth order of the LETGS, combined with the first order MEG count rate, is shown in \cite{Jellelightcurve2004A&A} Figure 1. We also show the light curve in our Fig.~\ref{fig:EWfit}. Here we summarize the characteristics of the light curve. The light curve is composed of the MEG and LETGS count rates over a span of $\sim625$ ks corresponding to the spectra we used in Table~\ref{tab:spectra}. In the first 400 ks of the observation, there is a gradual rise. Then it is followed by a decay with a similar time scale up to t = 520 ks. After that time, the light curve remains approximately flat for $\sim100$ ks till the end of the observation. Two data gaps are caused by the perigee passage of Chandra. The first gap is between the MEG and the LETGS observations, and the second gap occurs during the maximum flux.

To investigate potential spectral variations, we divide the entire observational period into five phases, as illustrated in Fig.~\ref{fig:EWfit}. The division aims to ensure sufficient count statistics in each phase while effectively capturing the variability. As listed in Table 1, the MEG and LETGa phases correspond to the pre-flare, LETGb represents the rise, LETGc the peak, and LETGd the post-flare period.

\renewcommand{\arraystretch}{1.4}
\begin{table}
\centering
\caption{The equivalent width (EW) values of \ion{Mg}{XII} Ly$\alpha$ and \ion{O}{VIII} Ly$\beta$ for the MEG and four split LETGS spectra.}
\begin{tabular}{lccc}
\hline\hline
Spectra & \ion{Mg}{XII} Ly$\alpha$ & \ion{O}{VIII} Ly$\beta$ &Time (ks)\\ \hline
MEG     &10.2$\pm$1.3    &39.9$\pm$3.2     & 100.6\\ \hline
LETGSa  & 19.7$\pm$5.9    &20.0$\pm$7.5 & 262.8    \\ \hline
LETGSb  &21.2$\pm$3.8    &23.5$\pm$5.6 & 334.2    \\ \hline
LETGSc  &9.6$\pm$4.0    &20.7$\pm$4.8  & 478.0   \\ \hline
LETGSd  &12.4$\pm$2.1    &17.6$\pm$4.8 & 574.3    \\ \hline
\end{tabular}
\tablefoot{The unit of EW is m$\AA$. The column time corresponds to the middle time of the spectrum, counted from the start of the first observation.}
\label{tab:EWvalue}
\end{table}

\subsection{Absorption lines variation}
\label{sec:Absorbption lines variation}
The flux increased during the observation from the MEG period to the LETGS observation (see Fig.~\ref{fig:EWfit} orange points). The average flux level differed by 30\% between both observations on a timescale of about 300 ks. We search for a possible response of the WAs to the change in ionizing flux. We first checked the results reported by \cite{steenbruggelinesearch2005A&A}. They mentioned that there may be a small enhancement of the \ion{Si}{XIV} and \ion{Mg}{XII} Ly$\alpha$ and the \ion {Si}{XIII} resonance line, corresponding to a $\sim50\% $ increase in ionic column density in the LETGS observation. However, the \ion{Si}{XIV} and \ion{Si}{XIII} resonance lines are too weak and sometimes hard to measure in both the MEG spectrum and the separate LETGS spectra. We only found the variation of \ion{Mg}{XII} Ly$\alpha$. Instead of silicon absorption lines, we found a significant variation in the strongest absorption line \ion{O}{VIII} Ly$\beta$. The EW of the absorption features was determined by fitting the local spectrum within 0.4 $\AA$ of the two absorption lines. Our fitting model consisted of a power-law continuum and a single absorption component represented by a negative Gaussian profile. The resulting best-fit profiles are shown in Fig.~\ref{fig:lineprofile}. As demonstrated in Fig.~\ref{fig:EWfit} and detailed in Table~\ref{tab:EWvalue}, the EWs of both the \ion{Mg}{XII} Ly$\alpha$ and \ion{O}{VIII} Ly$\beta$ lines exhibit time variations. The EW uncertainties reported in Table~\ref{tab:EWvalue} correspond to the 1$\sigma$ confidence level. Given that neither the \ion{Mg}{XII} Ly$\alpha$ nor the \ion{O}{VIII} Ly$\beta$ lines are saturated at their peak EW, these EW variations can be directly and linearly translated into changes in the ionic column density.

\begin{table}
\centering
\caption{Fits to the time-resolved equivalent widths with a constant value EW$\_$C (m$\AA$).}
\begin{tabular}{lcccc}
\hline\hline
\multicolumn{1}{l}{\multirow{2}{*}{Absorption lines}} & \multicolumn{2}{c}{MEG+LETGS}                                                        & \multicolumn{2}{c}{LETGS}                  \\ \cline{2-5} 
\multicolumn{1}{l}{}                                  & \multicolumn{1}{l}{EW$\_$C} & \multicolumn{1}{c}{$\chi^2$/dof} & EW$\_$C& $\chi^2$/dof \\ 
\hline
\ion{Mg}{XII} Ly$\alpha$                                        & 11.8                                  & 9.8/4                               
&14.1              & 6.3/3         \\
\ion{O}{VIII} Ly$\beta$                                       & 28.6                                  & 22.9/4                             & 20.3             & 0.7/3         \\ \hline
\end{tabular}
\label{tab:EWfitting}
\end{table}

To measure the response of the warm absorber to the ionization flux variations, we fitted the EW by a constant EW$\_$C. The fit results are shown in Fig.~\ref{fig:EWfit} and the best-fit EW$\_$C is listed in Table~\ref{tab:EWfitting}. We also investigated two cases: the MEG observation plus LETGS observations (blue lines) and only LETGS observations (purple lines).

\begin{figure*}
\centering
\includegraphics[width=\textwidth]{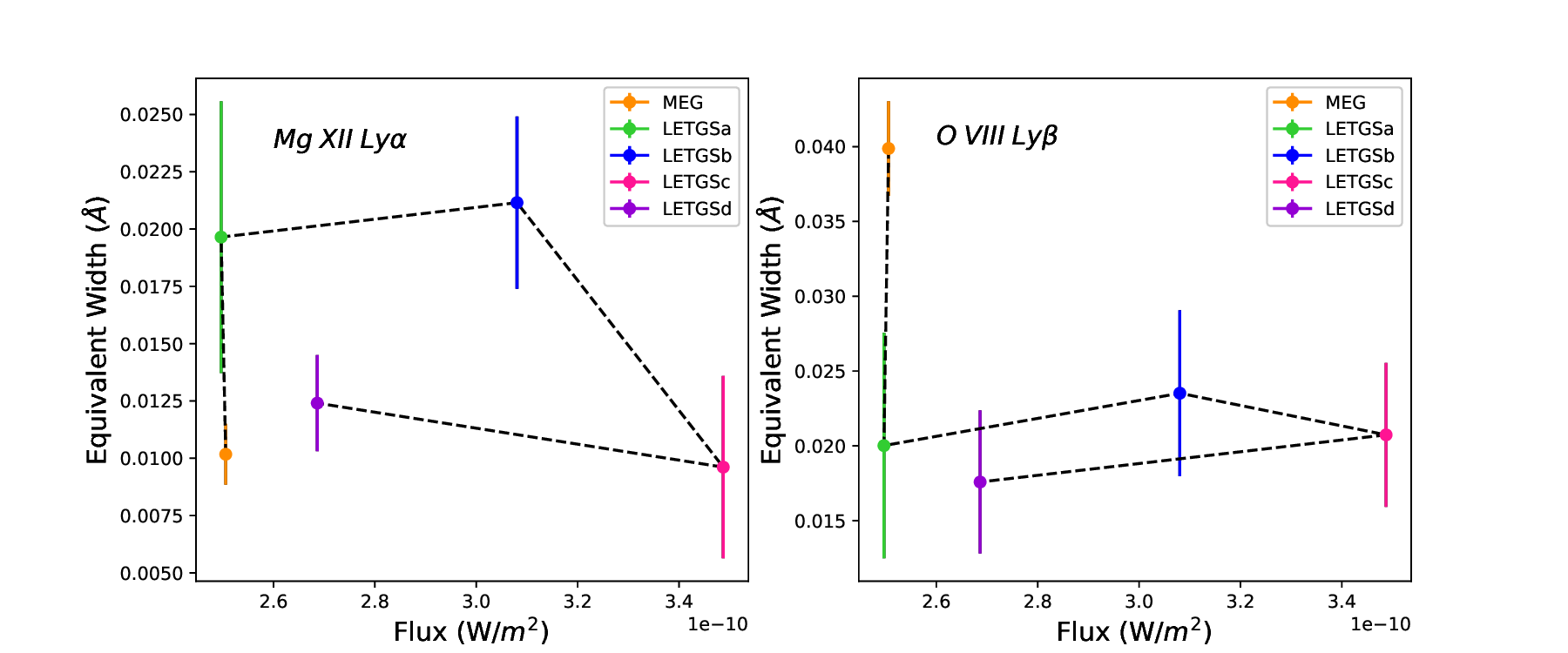}
\caption{The equivalent width of \ion{Mg}{XII} Ly$\alpha$ and \ion{O}{VIII} Ly$\beta$ lines, plotted as a function of the continuum flux in 1.5-24 $\AA$.}
\label{fig:EWflux}
\end{figure*}

As shown in Table~\ref{tab:EWfitting}, assuming a constant EW for \ion{Mg}{XII} Ly$\alpha$ can be rejected at a confidence level of $96\%$ when all data are combined, or at $90\%$ based on the LETGS data alone. The LETGb phase exhibits a high EW, exceeding the average by approximately 2.5 $\sigma$. Fig.~\ref{fig:EWflux} plots the EW against the flux in the 1.5–24 $\AA$ band. The \ion{Mg}{XII} Ly$\alpha$ EW generally increases with rising flux. However, in the peak and post-flare phases, the \ion{Mg}{XII} Ly$\alpha$ EW sharply drops. The significant variability provides an opportunity to explore the lower density limit.

Similarly to \ion{Mg}{XII} Ly$\alpha$, the EW of the \ion{O}{VIII} Ly$\beta$ absorbtion line exhibits significant variability at the $99.9\%$ confidence level. During the pre-flare phase observed with MEG, the line shows an excess corresponding to approximately 3.6$\sigma$ above the average. Meanwhile, the rising, peak, and post-flare phases show a constant \ion{O}{VIII} Ly$\beta$ line. The EW of \ion{O}{VIII} Ly$\beta$ exhibits opposite trends between the MEG and LETGS observations: As the light curve rises during the MEG and the beginning of the LETGS observation, the EW decreases (orange and green data points in Fig.~\ref{fig:EWflux} ) and then remains nearly constant for the remainder of the LETGS observation. The drop in EW for the \ion{O}{VIII} Ly$\beta$ absorption line between the MEG and LETGS observations suggests possible density constraints.

In total, there is almost 30$\%$ variability in the continuum flux for the most extreme cases. The EW of the LETGSa and LETGSb spectra for the \ion{Mg}{XII} Ly$\alpha$ absorption line is twice as high as those of the MEG and LETGSc–LETGSd spectra. The EW jump between LETGSb-LETGSc occurs during the peak of the flare in continuum flux. The EW of \ion{O}{VIII} Ly$\beta$ seems to be constant within the LETGS observation, but was much higher during the MEG observation. The EW value of \ion{O}{VIII} Ly$\beta$ decreased $\sim50\%$ between the MEG and the LETGSa observation.

\begin{table*}[]
\centering
\caption{Parameters of the six warm absorber components in NGC 5548 from \cite{Junjie2017A&A}.}
\begin{tabular}{lcccccc}
\hline\hline
Components & A & B & C & D & E & F \\ \hline
N$_\text{H}$ ($10^{24}$ m$^{-2}$)& 2.6$\pm$0.8  &6.9$\pm$0.9   &10.8$\pm$2.8   &13.4$\pm$2.1  & 25$\pm$13  & 52.0$\pm$8.5  \\ \hline
log$_{10}(\xi)$  &0.51$\pm$0.12&1.35$\pm$0.06  &2.03$\pm$0.04   & 2.22$\pm$0.03   & 2.47$\pm$0.13  & 2.83$\pm$0.03  \\ \hline
v$_\text{b}$ (km s$^{-1}$) &150$\pm$29&49$\pm$14   & 40$\pm$10  &67$\pm$17   & 6$\pm$5  & 115$\pm$29  \\ \hline
v$_\text{out}$ (km s$^{-1}$)& -557$\pm$37  & -547$\pm$35  & -1108$\pm$31  & -271$\pm$24  & -670$\pm$14  &-1122$\pm$34   \\ \hline
\ion{Mg}{XII} Ly$\alpha$ fraction        &  0.02 & 0.02  & 0.21  &  0.31 & 0.16 & 0.28  \\ \hline
\ion{O}{VIII} Ly$\beta$  fraction         &  0.05 & 0.25  & 0.21  & 0.25  & 0.09  & 0.15  \\ \hline
\end{tabular}
\tablefoot{The contribution fraction of the dominant components of \ion{Mg}{XII} Ly$\alpha$ and \ion{O}{VIII} Ly$\beta$ are listed at the last two rows.}
\label{tab:components}
\end{table*}

\subsection{Density constraint}
\label{sec:density constraint}
 We have found that the EW of the \ion{Mg}{XII} Ly$\alpha$ line decreased in the $\sim$144 ks between the LETGSb and LETGSc observations, and the \ion{O}{VIII} Ly$\beta$ line in the $\sim$162 ks between the MEG and LETGSa observations. This can be considered as the variability timescale. The warm absorber in NGC 5548 is composed of six distinct ionization components, labeled A to F in order of increasing ionization (\citealt{Junjie2017A&A}). Each component is represented by a PION component in SPEX with column density $N_\mathrm{H}$, ionization parameter $\xi$, outflow velocity $v_\mathrm{out}$, and turbulent velocities $v_\mathrm{b}$ (see Table~\ref{tab:components}). Based on the assumption that the variability in the \ion{Mg}{XII} Ly$\alpha$ and \ion{O}{VIII} Ly$\beta$ lines in NGC 5548 is driven by the response of warm absorbers to changes in the observed continuum flux. Here we investigate the specific component responsible for the variation and derive its density from the observed timescale.

 \begin{figure*}
\centering
\includegraphics[width=\textwidth]{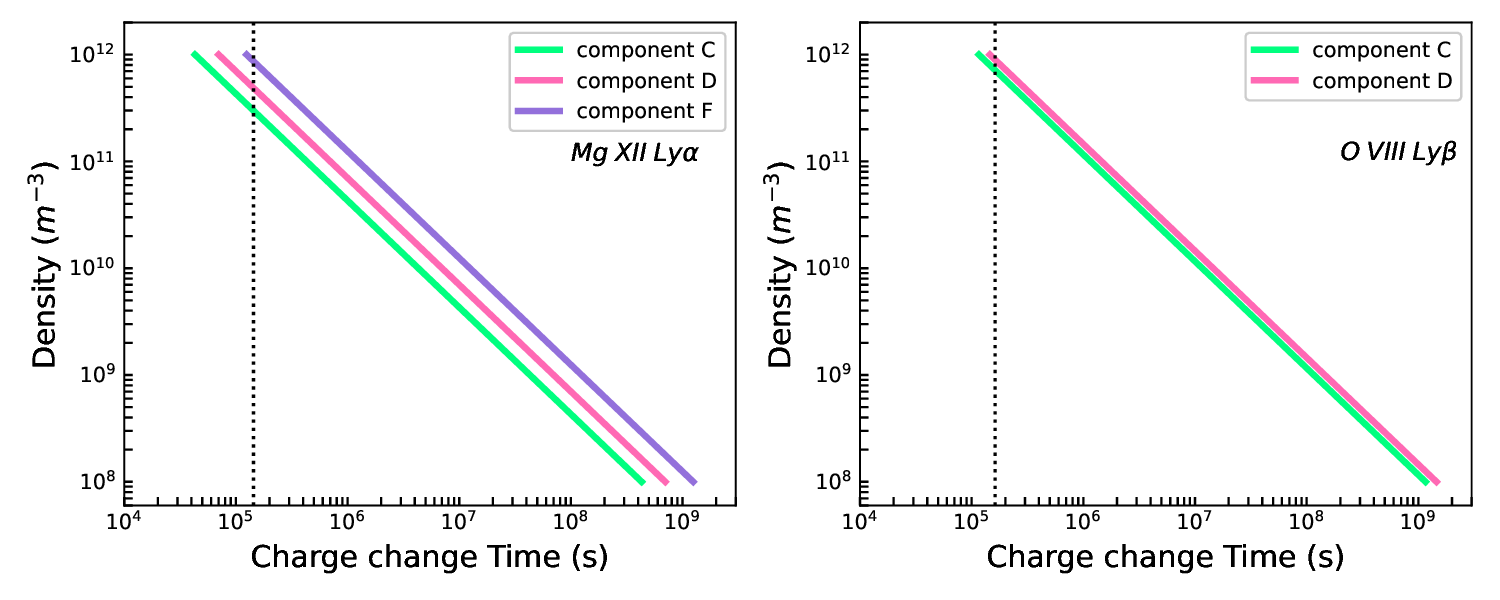}
\caption{The relationship between density and charge change time calculated by the PION models for \ion{Mg}{XII} Ly$\alpha$ and \ion{O}{VIII} Ly$\beta$ in their dominant components. The dotted line marked the variability timescale (144 ks for \ion{Mg}{XII} Ly$\alpha$ and 162 ks for \ion{O}{VIII} Ly$\beta$).}
\label{fig:den_time}
\end{figure*}
 
We consider components with a contribution fraction greater than 0.2 for each element here as significance (shown in Table~\ref{tab:components}). Components C, D, and F contribute significantly to the EW of the \ion{Mg}{XII} absorption line (0.21, 0.31, and 0.28 shown in Table~\ref{tab:components}, respectively). Similarly, components B, C, and D provide significant contributions of EW to the \ion{O}{VIII} absorption line (0.25, 0.21, and 0.25, respectively). Components C, D, and F exhibit variability timescales of approximately 2-4 days, while component B has a variability timescale that exceeds 200 days (\citealt{Ebrero2016A&A}), which is substantially longer than the variability observed in this study. The EW values have changed by 50$\%$ for both the \ion{Mg}{XII} Ly$\alpha$ and \ion{O}{VIII} Ly$\beta$ absorption lines. Changes in only one component cannot account for such significant variability, likely from multiple components. We also calculated the average charge states of oxygen and magnesium in these components. The average charge states of oxygen are 8.68 and 8.81 for components C and D, respectively. The average charge states of magnesium are 11.92, 12.28, and 12.91 in components C, D, and F, respectively. Therefore, we cannot specify which components have changed or how many components contribute to the observed EW variability. It is likely that one or more components C, D, and F collectively contribute to the EW variation of the \ion{Mg}{XII} Ly$\alpha$ line, with components C and D also influencing the EW changes of \ion{O}{VIII} Ly$\beta$. The observed decrease in magnesium EW at the flare peak can be explained by a balancing effect: as the flux increases, the EW contributions from component C slightly rise, while those from components D and F decline. In contrast, the EW pattern for O is simpler, as both components C and D consistently cause the EW to decrease as the flux increases.

Therefore, the differing patterns observed in the \ion{Mg}{XII} Ly$\alpha$ and \ion{O}{VIII} Ly$\beta$ variations, as shown in Fig.~\ref{fig:EWfit}, can be attributed to multiple warm absorber components contributing to the \ion{Mg}{XII} Ly$\alpha$ and \ion{O}{VIII} Ly$\beta$ lines, and to their intrinsic different responses to flux change due to different ionization degrees.

 According to \cite{Junjie2017A&A}, components C, D and F have ionization parameters of log$\xi\sim$2.03, 2.22 and 2.83, respectively (see Table~\ref{tab:components}). Assuming that the changes in the WA of the MEG and LETGS observations occur only due to ionization/recombination processes (i.e., outflow velocity $v_\mathrm{out}$ and column density $N_\mathrm{H}$ remain the same), the response to flux variations on different timescales is determined by the different densities of the absorbing gas components. We then used the PION model in SPEX, which outputs the recombination time, the charge change time ($t_{c}$), and the concentration as a function of the density. These are calculated from the ionization and recombination rates ($\alpha$). Thus, we can derive the density ($n_{e}$) constraint by applying the variability timescale as the charge change time to this program ($t_{c}\sim1/(n_{e}\alpha)$). We used the 144 and 162 ks time scales mentioned before for the \ion{Mg}{XII} Ly$\alpha$ and \ion{O}{VIII} Ly$\beta$ absorption lines as upper limits for the charge variation time scale with the results of the above program. The relationship between density and charge change time for \ion{Mg}{XII} Ly$\alpha$ and \ion{O}{VIII} Ly$\beta$, calculated by the PION models, are plotted in Fig.~\ref{fig:den_time}. The input parameters for the PION model were described above. The plot covers a density range of $10^{8}-10^{12}$m$^{-3}$ for each component, which is the most likely range where we could observe the variability. If the change originates from warm absorber component C, we derive a lower limit of the density of the warm absorber C from the \ion{Mg}{XII} Ly$\alpha$ of 2.7$\times10^{11}\text{m}^{-3}$ and from the \ion{O}{VIII} Ly$\beta$ of 7.2$\times10^{11}\mathrm{m}^{-3}$, If the change is due to the warm absorber D, the lower limit of the density from \ion{Mg}{XII} Ly$\alpha$ is 4.9$\times10^{11}\text{m}^{-3}$ and from \ion{O}{VIII} Ly$\beta$ is 9.0$\times10^{11}\mathrm{m}^{-3}$. If the change originates from the warm absorber F, we can only obtain the lower limit from \ion{Mg}{XII} Ly$\alpha$, which is 8.7$\times10^{11}\mathrm{m}^{-3}$. Thus, we adopt the most conservative constraints as our final density lower limits: the constraint from \ion{O}{VIII} Ly$\beta$ as the lower limit of density 7.2$\times10^{11}\mathrm{m}^{-3}$ if component C is responsible, 9.0$\times10^{11}\mathrm{m}^{-3}$ from \ion{O}{VIII} Ly$\beta$ if component D is responsible, and 8.7$\times10^{11}\mathrm{m}^{-3}$ from \ion{Mg}{XII} Ly$\alpha$ if component F is responsible.

\section{Discussion}
\label{sec:discussion}
\subsection{Comparison with previous results}
The 2002 Chandra grating data of NGC 5548 are among the best datasets for studying warm absorber variability because the high quality enables phase-resolved spectral analysis during a significant source flare. Previous work also discussed the variability of the warm absorbers. \cite{steenbruggelinesearch2005A&A} discussed the short time variability during the 2002 observation, but did not detect significant variations of the warm absorber as a response to the continuum flare occurring during the LETGS observation. \cite{Ebrero2016A&A} has shown the variation in the ionization parameter on different time scales for each warm absorber component in NGC 5548. They found that component C, D, and F started showing marginal evidence of variability at 2-4 days, which is consistent with our variation timescales ($\sim$162 ks), and derived the lower limits of the density of component C, D, and F using the time-average spectra during LETGS observation. As we split the LETGS observation into four parts, compared to \cite{steenbruggelinesearch2005A&A} using the average LETGS spectra, we were able to find the variations between the MEG and LETGS spectra and within the LETGS observation. This may indicate that some variations of the features were lost due to averaging the spectrum. We derive lower limits for the density of variable components using the shortest variability timescale observed so far for NGC 5548. Our derived lower limits of 7.2$\times10^{11}\mathrm{m}^{-3}$, 9.0$\times10^{11}\mathrm{m}^{-3}$ and 8.7$\times10^{11}\mathrm{m}^{-3}$ in components C, D, F are all nearly 5 times higher than the upper bounds of the reported ranges from \cite{Ebrero2016A&A}, representing a significantly more stringent constraint on the warm absorber density.

\subsection{The distance of the warm absorber}
The precise locations and mechanisms responsible for the launching of the warm absorbers remain outstanding questions with regard to the interpretation of the observations. Distances can be constrained indirectly through measurements of the ionization parameter, ionizing luminosity, and density through the definition of the ionization parameter (\citealp{Ionizationparameter11969ApJ,Ionizationparameter21995ApJ}):
   \begin{equation}
      \xi= \frac{L_{\mathrm{ion}}}{n_{\mathrm{H}}\times r^{\text{2}}} \,,
   \end{equation}
where L$_{\text{ion}}$ is the 1–1000 $\mathrm{Ryd}$ (or 13.6 eV–13.6 keV) band luminosity of the ionizing source, $n_{\mathrm{H}}$ the hydrogen number density of the ionized plasma, and $\mathrm{r}$ the distance of the plasma with respect to the ionizing source. Therefore, we can use the lower limits on the density $n_{\mathrm{H}}$ calculated in Section~\ref{sec:density constraint} to constrain the locations of the variable warm absorber components with respect to the central ionizing source. L$_{\mathrm{ion}}$= 1.97$\times10^{37}\mathrm{W}$ was calculated by integrating the template SED from \cite{steenbruggelinesearch2005A&A}. Using this value along with the $\xi$ value for components C, D and F (\citealt{Junjie2017A&A}) and assuming that these components are responsible for the observed EW variability, we derive the upper limits on the distance of the warm absorber components C of $\sim0.5$ pc and $\sim0.4$ pc for component D and $\sim0.2$ pc for component F. These values are up to a factor of two lower than the upper limit reported by \cite{Ebrero2016A&A}.

\section{Conclusions}
We have re-analyzed the archival observations of NGC 5548 taken with Chandra (LETGS and HETGS) during January 2002. We split the LETGS spectra into four parts to study the possible variation during the LETGS observation. We found that the \ion{Mg}{XII} Ly$\alpha$ and \ion{O}{VIII} Ly$\beta$ have significant variations as a response to the continuum flare occurring between the HETG and LETGS observations. Assuming that the observed changes in WA between different archival observations are driven purely by ionization and recombination processes, and using the variability timescales, we are able to constrain the lower limit on the density of the variable WA components to be $\sim$7.2-9.0$\times10^{11}\text{m}^{-3}$. Furthermore, lower limits on density can be used to estimate upper limits on the location of the WA. We find that the variable WA components are located within $\sim$0.2-0.5 pc from the central ionizing source. 

\begin{acknowledgements}
The scientific results reported in this article are based on observations made by the Chandra X-ray observatory. K. Zhao thanks for finical support from the Chinese Scholarship Council (CSC) and Leiden University/Leiden Observatory. SRON is supported financially by NWO, the Netherlands Organization for Scientific Research.
\end{acknowledgements}

\bibliographystyle{aa} 
\bibliography{aa53743-25.bib} 
\end{document}